\documentclass[12pt]{article}

\pdfoutput=1

\usepackage{graphics}
\usepackage{amssymb}
\usepackage{amsmath}

\newcommand{\rf}[1]{(\ref{#1})}
\newcommand{\beq}{\begin{equation}}

\newcommand{\eeq}{\end{equation}}
\newcommand{\bea}{\begin{eqnarray}}
\newcommand{\eea}{\end{eqnarray}}

\newcommand{\e}{\mbox{e}}

\newcommand{\gam}{\gamma}
\newcommand{\Gam}{\Gamma}

\newcommand{\Lam}{\Lambda}
\newcommand{\bt}{\beta}
\newcommand{\al}{\alpha}

\renewcommand{\th}{\theta}
\newcommand{\vep}{\varepsilon}     

\newcommand{\kp}{\kappa}

\newcommand{\oh}{\frac{1}{2}}
\newcommand{\oq}{\frac{1}{4}}

\newcommand{\prt}{\partial}

\newcommand{\cP}{{\cal P}}

\newcommand{\tx}{{\tilde{x}}}
\newcommand{\tG}{{\tilde{\Gam}}}
\newcommand{\tV}{{\tilde{V}}}

\newcommand{\R}{{\rm I\!R}}

\newcommand{\vecx}{\vec{x}}

\begin{document}

\begin{center}
${}$\\
\vspace{60pt}
{ \Large \bf Wilson loops in CDT quantum gravity}

\vspace{46pt}

{\sl J. Ambj\o rn}$\,^{a,b}$,
{\sl A. G\"orlich}$\,^{a,c}$,
{\sl J. Jurkiewicz}$\,^{c}$
and {\sl R. Loll}$\,^{b}$

\vspace{24pt}
{\footnotesize

$^a$~The Niels Bohr Institute, Copenhagen University\\
Blegdamsvej 17, DK-2100 Copenhagen \O , Denmark.\\
{email: ambjorn@nbi.dk, goerlich@nbi.dk}\\

\vspace{10pt}

$^b$~Institute for Mathematics, Astrophysics and Particle Physics (IMAPP)\\ 
Radboud University \\ 
Heyendaalseweg 135, 6525 AJ Nijmegen, The Netherlands.\\ 
{email: r.loll@science.ru.nl}\\

\vspace{10pt}

$^c$~Institute of Physics, Jagellonian University\\
Reymonta 4, PL 30-059 Krakow, Poland.\\
{email: jurkiewi@thrisc.if.uj.edu.pl}

\vspace{10pt}

}
\vspace{48pt}

\end{center}


\begin{center}
{\bf Abstract}
\end{center}

\noindent By explicit construction, we show that one can in a simple 
way introduce and measure gravitational holonomies and Wilson loops in lattice formulations 
of nonperturbative quantum gravity based on (Causal) Dynamical Triangulations. 
We use this set-up to investigate a class of Wilson line observables 
associated with the world line of a point particle coupled to quantum gravity, and
deduce from their expectation values that the underlying holonomies cover the group manifold
of SO(4) uniformly.

\vspace{12pt}
\noindent


\newpage

\section{Introduction}\label{intro}

Our quest for a theory of quantum gravity meets with numerous challenges. We not only have to define
the theory nonperturbatively, but must also ascertain that it actually exists and has desirable physical 
properties, including a well-defined classical limit. Many steps in this construction must be formulated
in terms of observables, which in a diffeomorphism-invariant theory are notoriously hard to come by. 
In this article, we focus on a particular class of observables, involving gravitational Wilson loops, and
a particular candidate theory of nonperturbative quantum gravity, Causal Dynamical Triangulations (CDT)
\cite{ajl,ajl1,agjl}. In this theory, the challenges mentioned above pose themselves in very concrete terms and
can also be addressed concretely, including the use of powerful numerical methods.

To understand our analysis of Wilson loops, a comprehensive understanding of CDT quantum gravity will not 
be necessary\footnote{The interested reader may consult our overview and review articles \cite{cdtreviews,physrep}.}; 
we will confine ourselves to a brief description of the approach, and in later sections
give some details of the geometric set-up, to the extent they are needed. 
In a nutshell, CDT is a covariant, quantum field-theoretic lattice formulation of gravity, where the nonperturbative
sum over spacetime geometries is realized in terms of piecewise flat four-geometries. They
are assembled from four-dimensional Lorentzian building blocks in such a way that
only causally well-behaved spacetime histories are included in the path integral. 

To perform the actual sum over these histories one must rotate them to Euclidean signature.
It is important to understand that not all Euclidean triangulations lie in the image of the
Wick rotation map, but only those in a subset, which carry a memory of the causal properties of 
their Lorentzian origin, most notably, the absence of ``baby universes" (topology changes of
spatial slices). 
The triangular building blocks or four-simplices are characterized by their side length $a$,
which plays the role of a UV cut-off. The continuum limit
of the regularized path integral involves a limit 
$a \to 0$, possibly accompanied by a readjustment of the bare 
coupling constants, such that physics stays invariant.  
In recent work \cite{RG} we demonstrated explicitly how a renormalization group flow is implemented in 
CDT quantum gravity, despite the absence of a background metric and the absence of any 
obvious correlation length. Apart from being a rather remarkable result, our analysis highlighted
the need for further observables to provide independent checks on our condition of keeping 
physics constant while altering the renormalization group scale. 
 
Only a few observables are known in CDT quantum gravity and have been investigated
quantitatively, including the volume profile of the dynamically generated quantum universe
\cite{agjl,volprofile}, as well as its Hausdorff and spectral dimensions \cite{ajl1,spectral}. Note
that all of them involve measurements of lengths and volumes. This is in contrast with
the classical continuum theory, where one describes the nontrivial, local structure of spacetime
in terms of its curvature, which is a function of the {\it derivatives} of the spacetime metric $g_{\mu\nu}(x)$.
A key question we would like to answer is 
whether there is a meaningful notion of ``curvature" or ``quantum curvature" in nonperturbative
quantum gravity, which on the Planck scale is well-defined and yields finite values, and on macroscopic
scales goes over into one of the standard curvatures of general relativity.

In Regge Calculus \cite{reggecalc} and Dynamical Triangulations\footnote{Lattice gravity 
in terms of Dynamical Triangulations (DT) 
is the purely Euclidean precursor of Causal Dynamical Triangulations, see, for example, \cite{aj}. 
Our theoretical considerations about Wilson loops presented below, up to and including Sec.\ \ref{impl}, 
are also applicable to DT.}
there is a simple, discretely defined expression for the local
scalar curvature in terms of deficit angles, which we will review in Sec.\ \ref{holosec} below,
but unfortunately it becomes singular in a na\"ive continuum limit. 
This is not at all surprising since the continuum definition of the
curvature involves second derivatives of the metric and a typical field configuration $g_{\mu\nu}(x)$ in 
the path integral is not expected to have well-behaved derivatives. 

A main motivation for considering Wilson loops to try to define some coarse-grained
measure of curvature comes from gauge field theory. Here one can construct a nonlocal,
gauge-invariant observable by taking the (trace of the) path-ordered exponential of the 
gauge potential $A_\mu(x)$ along a closed curve $\gamma$, to obtain the so-called
Wilson loop \cite{wilson}
\beq
\label{wilsongft}
W_\gamma (A)= {\rm Tr}\, {\cal P}\exp \oint_\gamma A,
\eeq
with $\cal P$ denoting path ordering. The relation with the local curvature tensor $F_{\mu\nu}(x)$
is exhibited by expanding the path-ordered exponential (the holonomy) around an infinitesimal
square loop of side length $\epsilon$ in the $\mu\nu$-plane, yielding
\beq
\label{pathord}
{\cal P}\exp \oint_{\gamma_{[\mu\nu]}}\!\!\!\!\!\!\! A \, =\, {\bf 1} + g\, F_{\mu\nu}^a X_a\, \epsilon^2 +O(\epsilon^3),
\eeq
where $X_a$ are the generators of the Lie algebra of the gauge group and $g$ denotes the coupling
constant. Moreover, the scaling behaviour of large 
Wilson loops provides a test for whether the theory is confining. 
Wilson loop observables are robust in the sense that they have a natural representation
in terms of lattice variables in lattice gauge theory and have been used successfully in
numerical studies.  

In gravity, one can use the metric-compatible Levi-Civita connection $\Gamma^\lambda_{\mu\nu}(x)$ to construct
holono\-mies and gravitational analogues of Wilson loops, as we will describe in more detail in
Sec.\ \ref{holosec} below. 
The path-ordered exponential of $\Gamma$ along a path defines a notion of parallel transport of tangent 
vectors, and all physical information contained in the Riemann curvature tensor $R^\kappa_{\lambda\mu\nu}(x)$ 
can be retrieved from suitable infinitesimal holonomies, analogous to the situation in gauge theory 
captured by eq.\ (\ref{pathord}). However, Wilson loops are not diffeomorphism-invariant, unless the
underlying loops are defined in physical terms. Of course, this does not mean that one cannot construct
quantum observables that depend on holonomies or Wilson loops and {\it are} diffeomorphism-invariant.

Gravitational Wilson loops on spacetime have been
little studied, with the exception of work in perturbative quantum gravity \cite{modanese} and 
in the context of the search for a nonabelian Stokes' theorem \cite{diakonov}. The story is
different in canonical quantum gravity, where holonomies along {\it spatial} curves
play a prominent role in Loop Quantum Gravity \cite{lqg}. This approach differs radically
from perturbative quantum gravity where the dynamical variables are
local fields like the metric $g_{\mu\nu}(x)$. Instead, in loop quantum gravity nonlocal holonomies are taken as 
part of a set of fundamental variables in terms of which the entire quantum dynamics should
be expressed. In the quantum theory they are promoted to finite operators, which are
assumed to not need any renormalization. This is different from ordinary gauge theory, where the
expectation values of Wilson loops need to be renormalized. 
 
In this article we consider quantum gravity in the CDT formulation. 
Despite being nonperturbative, it is nevertheless an ordinary quantum field-theoretical framework. 
In order to extract physical information from suitable loop averages
when the lattice cut-off is taken to zero, we therefore expect that observables involving
Wilson loops will require renormalization. 

Motivated by the fact that Wilson loops  -- at least infinitesimal ones -- encode retrievable curvature information,
and encouraged by their success as observables in nonperturbative QCD, our ultimate goal 
is to construct and measure quantum curvature observables in nonperturbative quantum gravity based
on holonomies or Wilson loops. As explained earlier, they should also provide us with a notion of
averaging or coarse-graining\footnote{We note in passing that the analogous averaging problem
in classical general relativity has not been resolved (see, for example, \cite{classaver}).}, 
to allow for a comparison with ordinary macroscopic curvature in a semiclassical limit.  
We do not know a priori whether such observables exist, and we are not aware of an explicit construction
in any approach to nonperturbative quantum gravity.\footnote{Theoretical arguments were put forward in 
a different formulation of lattice gravity based on
Regge calculus, promoting large Wilson loops as carriers of 
nonperturbative information \cite{hamber}. Although sympathetic to the aim, we are unable to follow the
technical claims in \cite{hamber} or to understand how the construction can be implemented meaningfully in a 
nonperturbative context.} The results derived in this paper, involving both theoretical considerations and
numerical simulations in four dimensions, hopefully present a step in the direction of our main goal, as well as
demonstrating that CDT quantum gravity as a framework is perfectly suited to studying observables of
Wilson loop type. 
 
In what follows, we begin by reviewing holonomies in continuum gravity (Sec.\ \ref{holosec}), 
as well as their counterparts
in piecewise flat spaces and, more specifically, in dynamical triangulations (Sec.\ \ref{holodt}).
In Sec.\ \ref{invarhol}, we introduce the invariant angles characterizing a general SO(4)-holonomy, and derive an
explicit expression for an associated distribution of their possible values on the group manifold. 
A convenient choice of coordinate frames on the four-simplices of the triangulations is introduced
in Sec.\ \ref{impl}, as well as two different ways to compute the holonomies of closed lattice loops. 
In Sec.\ \ref{wilsonlines} the discussion focusses on a specific class of Wilson loops, associated with the
world line of a point particle, and their concrete implementation in the full, nonperturbative CDT path
integral. After a brief description
of the Monte Carlo simulation of the combined gravity-particle system, 
Sec.\ \ref{measure} contains our main computational result, the measured
distribution of the invariant angles for the class of Wilson lines considered. We conclude in Sec.\ \ref{discussion}
with a discussion and outlook.

\section{Holonomies in gravity}
\label{holosec}

The Levi-Civita connection $\Gam^\mu_{\nu\kp}(x)$ of a Riemannian manifold $M$ with metric $g_{\mu\nu}(x)$ defines a notion
of parallel transport of a vector $V^\mu$ along a curve $\gamma^\mu(\lambda)$. Transporting $V^\mu$ along the curve between
parameter values $\lambda_{\rm i}$ and $\lambda_{\rm f}$ results in a general linear transformation of the vector, which is given in terms of 
the path-ordered integral,
\beq\label{0.1}
V^\mu(x_{\rm f}) = 
\left(\cP\, \e^{-\int_{\lambda_{\rm i}}^{\lambda_{\rm f}} \Gam_\kp \dot\gamma^\kp(\lambda)d\lambda}\right)^\mu{}_{\nu}\, 
V^\nu(x_{\rm i}),~~~~\Big(\Gam_\kp\Big)^\mu{}_{\nu} = \Gam^\mu_{\kp\nu},
\eeq
where $\cP$ denotes path-ordering, the dot indicates differentiation with respect to the path parameter $\lambda$, and
$x_{\rm i}=\gamma(\lambda_{\rm i})$ and $x_{\rm f}=\gamma(\lambda_{\rm f})$ are the initial and final points of the path in $M$.

Under a coordinate transformation $x\to \tx(x)$, with $M^\mu{}_\nu(x)= \frac{\prt \tx^{\mu}(x)}{\prt x^{\nu}}$,
the path-ordered integral transforms nontrivially at its two endpoints $x_{\rm i}$ and $x_{\rm f}$, 
\beq\label{0.2}
\left(\cP\, \e^{-\int_{\lambda_{\rm i}}^{\lambda_{\rm f}} \tG_\kp \dot{\tilde\gamma}^\kp(\lambda)d\lambda}\right)^\mu{}_\nu =
M^\mu{}_\al(x_{\rm f})
\left(\cP\, \e^{-\,\int_{\lambda_{\rm i}}^{\lambda_{\rm f}} \Gam_\kp \dot{\gamma}^\kp(\lambda)d\lambda}\right)^\al{}_\bt 
(M^{-1}(x_{\rm i}))^\bt{}_\nu,
\eeq
in accordance with the transformation behaviour of vectors under coordinate transformations, namely,
\beq\label{0.3}
\tV^\mu(\tx) = M^\mu_{~\nu}(x) \, V^\nu(x).
\eeq 
Before turning to the case of a piecewise flat manifold, let us look at the concrete construction of
the path-ordered product $\cP \exp (-\int_\gamma \Gamma)$ for a given curve $\gamma(\lambda)$. 
In general, $\gamma(\lambda)$ will pass through several coordinate patches $U_k$, $k=1,\ldots,n$, with 
corresponding coordinates $x^\mu_k$. 

Let us consider the simplest situation of this kind, where the initial
point $x_{\rm i}$ of the curve lies in an open neighbourhood $U_0$ and the final point $x_{\rm f}$ in an open
neighbourhood $U_1$, such that $x_{\rm i}\notin U_1$, $x_{\rm f}\notin U_0$ and the intersection
$U_0\cap U_1$ is not empty. To perform the path integration along $\gamma$ of the connection
$\Gamma$, an intermediate point $\gamma(\lambda_{\rm mid})=x_{\rm mid}$ must be chosen in the overlap region $U_0\cap U_1$,
and the integration performed in two pieces: from $\lambda_{\rm i}$ to $\lambda_{\rm mid}$ over the connection $\Gamma_0(x_0^\mu)$ in
the coordinates $x_0^\mu$ of patch $U_0$, and subsequently over the connection $\Gamma_1(x_1^\mu)$ in terms
of the coordinates $x_1^\mu$ of $U_1$. In addition, to account for the change of coordinate system, 
a matrix $M(x_{\rm mid})=\frac{\prt x_1}{\prt x_0}|_{x_{\rm mid}}$ has to be inserted at the midpoint, leading to a combined expression
schematically given by
\beq\label{pathintcomb}
(\cP\, \e^{-\,\int_{x_{\rm mid}}^{x_{\rm f}} \Gam_1})^\mu{}_\nu 
\; M(x_{\rm mid})^\nu{}_\lambda 
(\cP\, \e^{-\,\int_{x_{\rm i}}^{x_{\rm mid}} \Gam_0 })^\lambda{}_\kappa .
\eeq
Using the transformation law (\ref{0.2}), it is straightforward to show that the value of expression (\ref{pathintcomb}) is independent 
of the choice of midpoint $x_{\rm mid}\in U_0\cap U_1$. 

\begin{figure}[ht]
\centering
\scalebox{0.5}{\includegraphics{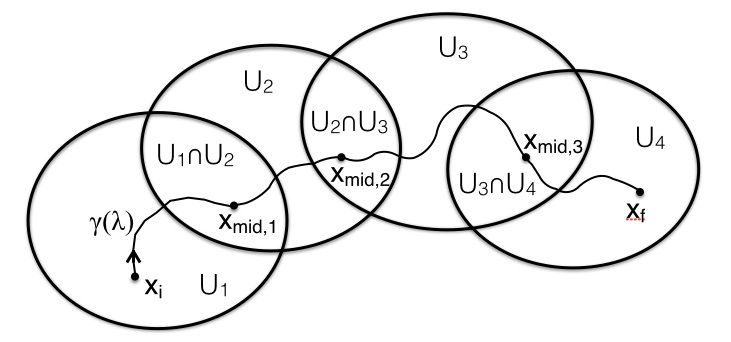}}
\caption[phase]{Starting at an initial point $x_{\rm i}$, a path $\gamma (\lambda)$ traverses a sequence of coordinate neighbourhoods $U_k$,
ending up at $x_{\rm f}$. A path-ordered integral along $\gamma$ of a connection is computed piecewise in every patch $(U_k,\{ x^\mu_k \} )$.
The switch from the $k^{\rm th}$ to the $(k+1)^{\rm st}$ coordinate system can happen at an arbitrary midpoint $x_{{\rm mid},k}\in U_k\cap
U_{k+1}$ and is associated with a matrix $M(x_{{\rm mid},k})$ as described in the text, c.f. eq.\ (\ref{0.5}).
}     
\label{neighbhd}
\end{figure}

If the path $\gamma$ runs through several coordinate neighbourhoods, the construction for the two patches 
just given can be reiterated, see Fig. \ref{neighbhd}. We are specifically interested in closed paths and
therefore will consider the situation where $\gamma$ starts at $x_{\rm i}$ in neighbourhood $(U_0,\{ x_0^\mu \} )$, 
passes through a sequence of $n$ neighbourhoods $( U_k,\{ x^\mu_k\})$, $k=1,\dots ,n$, via their non-empty
intersections $U_k\cap U_{k-1}\not= \emptyset$, and
finally into $U_{n+1}\equiv U_0$, where it ends up at the same point $x_{\rm f}\equiv x_{\rm i}$ it started from.
The path-ordered integral or holonomy associated with the oriented 
loop $\gamma$ based at $x_{\rm i}$ is then represented by 
\bea\label{0.4}
\left(\cP  \e^{-\oint_\gamma \Gamma}\right)_{x_{\rm i}}\!\!\! & = & (\cP\, \e^{-\,\int^{x_{\rm f}}_{x_{{\rm mid},n+1}} \Gam_0 }) \, M(x_{{\rm mid},n+1})\cdot \nonumber \\
&& \prod_{k=1}^{n} \big( (\cP\, \e^{-\,\int_{x_{{\rm mid},k}}^{x_{{\rm mid},k+1}} \Gam_k })\, M(x_{{\rm mid},k})\big)
\; (\cP\, \e^{-\,\int_{x_{\rm i}}^{x_{{\rm mid},1}} \Gam_0 }),
\eea
where it is understood that the matrix multiplication is from right to left as the loop parameter $\lambda$ in $\gamma(\lambda)$ increases from
$\lambda_{\rm i}$ to $\lambda_{\rm f}$. While the path-ordered integral appearing in eq.\ (\ref{0.1}) is valued
in GL(4,$\R$), the holonomy matrix (\ref{0.4}) on an orientable manifold is valued in SO(4).
The transformation matrix from coordinates $\{ x_{k-1}^\mu \}$ to $\{ x_k^\mu \}$ is given by
\beq\label{0.5}
M(x_{{\rm mid},k})^\alpha{}_\beta =\bigg( \frac{\prt x_k^\alpha}{\prt x_{k-1}^\beta }\bigg) |_{x_{{\rm mid},k}}
\eeq
and is evaluated at the $k^{\rm th}$ midpoint $x_{{\rm mid},k}$, to be chosen freely along $\gamma$ in the overlap region $U_k\cap U_{k-1}$. 
The holonomy (\ref{0.4}) still depends on the initial or base point $x_{\rm i}$ and under a coordinate transformation $ x\rightarrow \tilde x (x)$
will transform according to \rf{0.2} as
\beq\label{0.6}
\left(\cP \e^{-\oint_\gamma  \tG }\right)_{\tilde x_{\rm i}} 
= M(x_{\rm i})\left(\cP \e^{-\oint_\gamma \Gam }\right)_{x_{\rm i}} M^{-1}(x_{\rm i}), \;\;\;\;\; {\rm with}\;\;
M^\mu{}_\nu(x)= \frac{\prt \tx^{\mu}(x)}{\prt x^{\nu}}.
\eeq
It follows that the conjugacy class of the holonomy matrix $\cP \exp (- \oint_\gamma \Gamma)$ is coordinate independent.
It is also easy to show that it does not depend on the starting point $x_{\rm i}$ chosen along the loop $\gamma (\lambda)$.
In this paper we will precisely study such ccordinate-independent conjugacy classes.

\section{Holonomies in Dynamical Triangulations}
\label{holodt}

Studying holonomies in the context of piecewise flat geometries simplifies the above discussion considerably, as we will see. 
The building blocks of the piecewise linear geometries used in dynamical triangulations 
are identical, equilateral\footnote{In {\it causal} dynamical triangulations one usually works with two different edge lengths, one
for time-like and one for space-like edges \cite{physrep}. In the present study we 
consider for simplicity the special case where after the Wick rotation of CDT the two edge lengths are identical, and each
triangulation therefore becomes equilateral.} four-simplices, 
which by assumption are everywhere flat on the inside, like the building blocks of Regge calculus \cite{regge}.
When these building blocks are glued together along identical boundary three-simplices or ``faces" 
to construct a four-dimensional piecewise flat manifold, curvature will generically 
appear in a singular fashion along the two-dimensional subsimplices of the triangulation, the triangles or ``hinges".

Recall that the geometry of a four-simplex is completely fixed by its edge lengths (in our case the single 
edge length $\ell$), and that the geometric properties of a four-geometry assembled from such simplices are encoded in 
the gluing data (how faces are identified pairwise), neither of which requires the introduction of
coordinates.
Indeed, an important strength of the nonperturbative path integral formulation of CDT comes from the fact that no 
coordinates have to be introduced, and that the path integration does not contain unphysical coordinate reparameterizations
or other parameter redundancies. 

In the present piece of work we are {\it not} going to change the way we perform the path integral, but in order to analyze 
particular quantum operators involving holonomies we will introduce coordinate systems on individual four-simplices.
We are in principle completely free how to do this. Since the final result will not
depend on these choices, it is convenient to use the same Euclidean flat coordinate system on every simplex.
(We will specify our particular choice later on.) Since this makes the metric constant, the connection $\Gamma$ vanishes 
everywhere on the four-simplex,
and its path-ordered integral along any curve $\gamma$ is the unit matrix, as long as $\gamma$ remains inside the simplex.

Our considerations about computing holonomies of loops passing through several coordinate patches apply 
to these simplicial manifolds as follows. 
Two neighbouring four-simplices $s_1$ and $s_2$ with associated flat Euclidean coordinate systems $\{ x_1^\mu\}$ 
and $\{ x_2^\mu \}$ always have a three-simplex (tetrahedron) $\sigma$ in common. Since there is no curvature associated
with (the interior of) this three-dimensional interface, combining it with the interiors of the two four-simplices results
in a single open set whose geometry is flat and constant. Calling $U_i$ the coordinate patch parameterising (the interior of)
four-simplex $s_i$, this implies that $(U_1,\{ x_1^\mu \} )$ can be continued to the interior of $s_2$ and/or 
$(U_2, \{ x_2^\mu \} )$ can be continued to the
interior of $s_1$ to create a non-empty overlap region $U_1\cap U_2$. Like in the smooth case above, this allows us
to associate a path-ordered integral with any curve passing from $s_1$ to $s_2$ by inserting a matrix $M=\frac{\partial x_2}
{\partial x_1}$ associated with the change of coordinates $x_1\rightarrow x_2(x_1)$ 
in between the piece of the path-ordered integral computed in $U_1$ and that computed in $U_2$. 
Since the coordinate systems are flat Euclidean and moreover are the same for all simplices, the transition matrix $M$ 
on the four-dimensional overlap region $U_1\cap U_2$ is {\it constant} and given by a four-dimensional rotation.
Reverting once again to a simplicial description, we may therefore simply associate the matrix $M$ with the entire (interior of the) 
three-dimensional interface $\sigma$ between $s_1$ and $s_2$. To capture this simple dependence, we will introduce
a new notation for the corresponding rotation matrix, namely,
\beq
M^\mu{}_\nu =\frac{\partial x_2^\mu}{\partial x_1^\nu}|_\sigma =: R(s_2,s_1)^\mu{}_\nu \in O(4),
\label{rotdef}
\eeq
with the implicit understanding that $R$ still depends on the coordinates $\{ x_1^\mu\}$ and $\{ x_2^\mu \}$.
The explicit form of $R(s_2,s_1)$ depends on the relative orientation of the two coordinate frames $(s_1,\{ x_1^\mu\})$ 
and $(s_2,\{ x_2^\mu\})$, and can be computed once the coordinate systems have been specified.
We conclude that the path-ordered integral along any path crossing from $s_1$ to
$s_2$ anywhere in the interior of $\sigma$ will pick up a factor of $R(s_2,s_1)$. 

Consider now a closed path $\gamma (\lambda)$ in a piecewise flat simplicial manifold $\cal T$, and assume that it does not pass through any
of the two-dimensional subsimplices of $\cal T$, thereby avoiding potential curvature singularities.
The holonomy along $\gamma$ is then given by the ordered product of the rotation matrices associated
with subsequent crossings of $\gamma$ from one four-simplex to the next. Since the connection inside the four-simplices vanishes,
we can restrict ourselves to a limited set of standardized closed paths without losing any holonomy information. 
For our present purposes it is convenient to use only loops consisting of straight segments between the centres of neighbouring four-simplices.
For the holonomy $R_L$ of such a loop $L$, which passes through a sequence $s_1$, $s_2$, $\dots$, $s_n$, $s_1$ of four-simplices,
formula \rf{0.4} reduces to a product of the corresponding rotation matrices,  
\beq\label{1.1}
R_L = R(s_1,s_n)R(s_n,s_{n-1})\cdots R(s_2,s_1).
\eeq
Like the general holonomy (\ref{0.4}), $R_L$ still has a residual coordinate dependence and
transforms non-trivially at its base point under a coordinate transformation 
$x\rightarrow \tilde x(x)$ on $s_1$.
Since we have already fixed the coordinate frames of four-simplices to be flat and Euclidean, such a coordinate transformation must be
a $O(4)$-rotation $\Lambda$, and $R_L$ will transform by conjugation accordingly,
\beq\label{1.2}
R_L \to \Lam R_L \Lam^T, \;\;\;\; \Lambda^\mu{}_\nu =\frac{\partial\tilde x^\mu (x)}{\partial x^\nu}.
\eeq
If we parallel-transport a vector $V$ in the tangent space to $(s_1,\{ x_1^\mu\})$ around the loop $L$,
it will undergo a four-dimensional rotation to a new tangent vector
\beq\label{1.3}
V_L = R_L V. 
\eeq
The angle $\theta_{V,V_L}$ between the original and the rotated vector, defined as
\beq
\theta_{V,V_L} := \arccos \Big( \frac{V\cdot V_L}{\sqrt{V\!\cdot\! V} \sqrt{V_L\!\cdot\! V_L} }\Big),
\eeq
is independent of $\Lambda$, since a $SO(4)$-rotation preserves scalar products $V\cdot W$ 
of vectors in $\R^4$.

\begin{figure}[t]
\centering
\scalebox{0.5}{\includegraphics{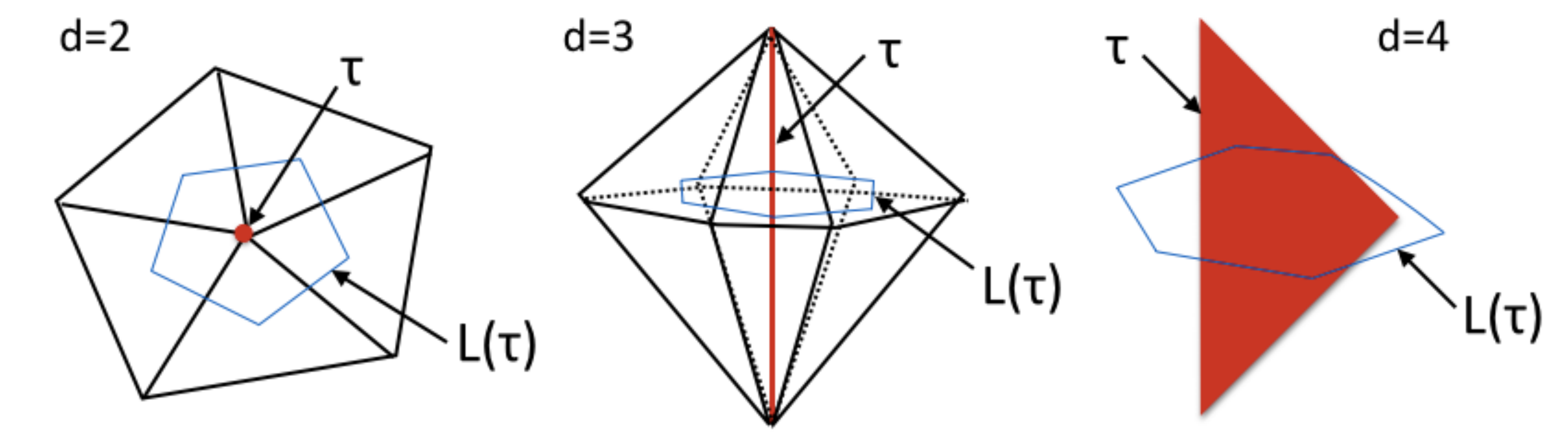}}
\caption[phase]{Local curvature associated with a hinge $\tau$ of codimension two in a $d$-dimensional simplicial manifold:
a vertex in $d\! =\! 2$, an edge in $d\! =\! 3$ and a triangle in $d\! =\! 4$. Parallel transport of a vector perpendicular to a hinge
$\tau$ along a minimal closed path $L(\tau )$ through the centres of the $d$-simplices which form the star of $\tau$ rotates the
vector by an angle equal to the deficit angle associated with $\tau$. (For ease of presentation, for $d\! =\! 4$ 
the four-dimensional star of $\tau$ is not shown.)}     
\label{hinges234}
\end{figure}

The set-up we have just introduced is closely related to how local curvature is described
in Regge calculus. Generalizing from four dimensions, the curvature of a $d$-dimensional simplicial manifold is
located at the subsimplices of dimension $d-2$. Given such a hinge $\tau$, which does not lie on the boundary of the
manifold, the curvature associated
to it can be thought of as the Gaussian curvature of a small two-dimensional surface perpendicular to $\tau$.
The surface is locally flat, with a conical singularity at the location where the hinge meets the surface. 
The magnitude of the curvature is quantified in terms of a deficit angle, which in a given simplicial manifold
can be extracted via parallel transport as we will explain below for the case $d=4$. 

To better understand the geometry of the situation, consider first the local confi\-guration of 
$d$-simplices sharing a hinge $\tau$. They form 
the so-called {\it star} of $\tau$, which topologically speaking is a $d$-dimensional ball. They also
form a circular neighbourhood around $\tau$, in the sense that we can construct a minimal closed
piecewise straight path $L(\tau)$ encircling $\tau$, which connects the centres of adjacent $d$-simplices 
in the star (see Fig.\ \ref{hinges234}).

Returning to the physically relevant case $d=4$, how will an arbitrary vector $V$ be affected by parallel transport 
along $L(\tau)$? 
Note first that $V$ will {\it not} be rotated at all if in the initial four-simplex $s_1$ it is parallel to the two-plane
spanned by the triangle $\tau$ (identifying the linear structure of the simplex and its tangent space). 
This can be easily understood as follows. In $s_1$, choose an
orthonormal coordinate system $\{ x_1^\mu \}$ such that the hinge $\tau$ lies in the plane spanned by $x_1^3$ and $x_1^4$,
say. Since $\tau$ is common to all four-simplices $s_k$ in the star of $\tau$, one can in each of them make the
same choice for the two coordinate axes $x_k^3$ and $x_k^4$ relative to $\tau$. In other words, an
arbitrary vector $V^\mu =(0,0,V^3,V^4)$ in $s_1$ will have exactly the same form in each of the $s_k$, independent of
the choice of the coordinates in the directions perpendicular to $\tau$. It follows that only components of $V$ perpendicular
to $\tau$ can be affected nontrivially by the holonomy matrix. Parallel transport around $\tau$ along the 
minimal loop $L(\tau)$ will therefore map a vector $V^\mu =(V^1,V^2,0,0)$ to some
$V_L^\mu =(V_L^1,V_L^2,0,0)$. The only $SO(4)$-transformations that can have this effect belong to the
one-parameter $SO(2)$-subgroup mapping the plane orthogonal to $\tau$ into itself. 
As a consequence, the rotation undergone by $V$ is characterized by a single angular parameter $\vep$, where 
\beq\label{1.4}
\cos \vep = \frac{V\cdot V_{L(\tau )}}{\sqrt{V\!\cdot\! V} \sqrt{V_{L(\tau)}\!\cdot\! V_{L(\tau)}} },\;\;\;\;{\rm for}\;\; V\perp \tau.
\eeq
We recognize $\vep$ as the deficit angle associated with the triangle $\tau$, as defined in Regge calculus,
and note its coordinate-independent character.   

To understand the range of the coordinate-invariant information that can be obtained by studying holonomies, 
let us recall some properties of the group $SO(4)$. A maximal torus of $SO(4)$ is given by the two-parameter set
of matrices 
\beq\label{1.5}
U(\theta_1,\theta_2) = 
\begin{pmatrix} 
\cos \th_1 & \sin \th_1 & 0&0 \\
-\sin \th_1 & \cos \th_1 & 0&0 \\
0 &0& \cos \th_2 & \sin \th_2  \\
0 & 0 & -\sin \th_2 & \cos \th_2 
\end{pmatrix},
\eeq
forming a maximal abelian subgroup $SO(2)\times SO(2)$. This implies that the rank of the group is 2. 
Moreover, given a compact connected Lie group $G$ -- like $SO(4)$ -- and a maximal torus $H\subset G$, 
each element $g\in G$ 
is conjugate to an element $h\in H$, that is, there is a $x\in G$ and a $h\in H$ such that $g=xhx^{-1}$ (see, 
for example, \cite{simon}). 

By the action of the holonomy $R_{L(\tau)}$ of a minimal loop $L(\tau)$ around a triangle $\tau$, 
with the coordinate choice made
above, a vector $V$ orthogonal to the $\tau$-plane according to relation \rf{1.4} will be rotated by  
an angle $\vep$, with corresponding holonomy matrix
\beq\label{1.6}
R_{L(\tau )} = \begin{pmatrix} 
\cos \vep & \sin \vep & 0&0 \\
-\sin \vep & \cos \vep & 0&0 \\
0 &0& 1 & 0  \\
0 & 0 & 0 & 1
\end{pmatrix}.
\eeq 
Note that in the CDT setting $\vep$ can only take one of a discrete set of values, because all
building blocks are identical and therefore all length and angular variables describing them come
in discrete units. 

It is important to realize that the matrix (\ref{1.6}) from the point of view of
$SO(4)$ corresponds to a particular type of rotation, a so-called ``simple rotation", which for
$SO(n)$-rotations on $\R^n$ is defined as a rotation that leaves a linear subspace of dimension 
$n-2$ fixed. The rotation matrix (\ref{1.6}), and any matrix obtained from it by conjugation, leaves a
two-plane through the origin fixed, and is therefore an example of a simple rotation in $SO(4)$. 
In dimensions $n\! =\! 2$ and $n\! =\! 3$ every rotation is simple, but this is no longer the case for $n\! =\! 4$,
where a generic rotation instead is characterized by {\it two} angles, and is conjugate to a matrix of
the form (\ref{1.5}), with both $\theta_1\not= 0$ and $\theta_2\not= 0$. Geometrically, such a ``double
rotation" consists of
two independent (and commuting) rotations in two two-planes which are mutually orthogonal and therefore 
share only one point, the origin, which is also the only point mapped into itself by this kind of rotation.

The distinction between a simple and a double rotation for a $SO(4)$-holonomy matrix is intrinsic and 
independent of coordinates. There is no coordinate transformation which will convert a double rotation
with $\theta_1\not= 0$ and $\theta_2\not= 0$ to a simple one. The fact that parallel transport around a single 
triangle $\tau$ in a simplicial four-dimensional manifold $\cal T$ results in a simple rotation has to do with the nature of the
curvature singularity located at $\tau$. Parallel transport around a more 
general loop in $\cal T$ will in general not lead to a simple rotation. The same holds for parallel transport 
around loops in a general curved continuum manifold, regardless of whether the loops are finite or 
infinitesimal.\footnote{The fact that {\it two} angles are
necessary to characterize holonomy matrices up to conjugation, as soon as one considers non-minimal
loops in a four-dimensional simplicial manifold $\cal T$, seems to have been overlooked by the 
authors of \cite{hamber}. This also holds when $\cal T$ is almost flat and holonomies do not
deviate much from the identity matrix.}

\section{Invariants from holonomies}
\label{invarhol}

Rather than operating with equivalence
classes of rotation matrices under conjugation, a convenient way of extracting the coordinate-invariant 
information of a holonomy matrix $R_L$ is to take its trace, Tr$(R_L)$. Especially in the context of
gauge field theory, where the path-ordered integral is taken over the local gauge connection, this
quantity is known as a Wilson loop. Because of the cyclic property of
the trace, it is invariant under conjugation,
\beq
{\rm Tr}(\Lambda R_L \Lambda^T)={\rm Tr}(R_L),\;\;\;\;  \Lambda\in SO(4),
\eeq
which means that for a generic holonomy matrix $R_L\in SO(4)$ we can from (\ref{1.5}) define the
invariant quantity
\beq
\label{trace}
t_1(R_L):=\frac{1}{2}\, {\rm Tr}(R_L)=\cos\theta_1 + \cos\theta_2.
\eeq
The fact that one can interchange the two $(2\times 2)$-blocks on the diagonal of the matrix (\ref{1.5}) by an
appropriate conjugation is reflected in the fact that the right-hand side of equation (\ref{trace})
is invariant under the exchange of $\theta_1$ and $\theta_2$. 
Assuming for the sake of definiteness that the angles $\theta_i$ take values in the interval $[0,2 \pi ]$,
we can by conjugation achieve that $(\theta_1,\theta_2)\mapsto (2\pi -\theta_1,2 \pi -\theta_2)$,
which likewise leaves (\ref{trace}) invariant. 

To extract information about both angles
$\theta_i$ separately, we can supplement expression (\ref{trace}) by a second invariant,
\beq
\label{tracesq}
t_2(R_L):=\frac{1}{4} \, {\rm Tr}(R_L^2)+1=\cos^2\theta_1+\cos^2\theta_2.
\eeq
If we fix the range of the angles to $[0,\pi [$ and require $\theta_1\leq \theta_2$, say,
the invariants $t_1$ and $t_2$ fix $(\theta_1,\theta_2)$ uniquely.

In the four-dimensional simulations we will extract this coordinate-invariant information by measuring 
$t_1$ and $t_2$ for a variety of closed curves, and on various ensembles of simplicial CDT geometries. 
As pointed out earlier, the $O(4)$-rotations which enter into the construction of these quantities are not
arbitrary but belong to a discrete set of possible rotations between neighbouring simplices. We will investigate
how this influences the measured invariants $t_i$. More specifically, we will extract from them the 
distribution of the angles $\theta_i$ and compare it to the distribution one
would obtain if the holonomy matrices were distributed uniformly over the group manifold of $SO(4)$. 

In order to do this, we need to derive the theoretical distribution of the $\theta_i$ on $SO(4)$. 
Recall that these angles were introduced in the context of the maximal torus (\ref{1.5}). They are
two of a total of six parameters needed to label points of $SO(4)$. By the theorem quoted earlier, every
$g\in SO(4)$ can be obtained from an element of the maximal torus by conjugation. An explicit way of
doing this, which introduces an explicit parametrization of the group manifold, is given by
\beq
\label{grouppara}
g(\theta_1,\theta_2,\omega_1,\omega_2,\varphi_1,\varphi_2):=
U(\varphi_1,\varphi_2)W(\omega_1,\omega_2)U(\theta_1,\theta_2)W(\omega_1,\omega_2)^T
U(\varphi_1,\varphi_2)^T,
\eeq
with $0\leq \theta_i <2\pi$, $0\leq \omega_i <\pi$ and $0\leq \varphi_i <2\pi$, and where 
\beq
\label{morerot}
W(\omega_1,\omega_2) = 
\begin{pmatrix} 
\cos \omega_2 & 0 & 0 & \sin \omega_2  \\
0 & \cos \omega_1 & \sin \omega_1 & 0 \\
0 & -\sin \omega_1 & \cos \omega_1 & 0  \\
-\sin \omega_2 & 0 & 0 &  \cos \omega_2 
\end{pmatrix}.
\eeq
A straightforward way to obtain the desired distribution $P(\theta_1,\theta_2)$ 
of the $\theta_i$ on $SO(4)$ is to compute
the Haar measure in terms of $\{ \chi_k \}=\{\theta_1,\theta_2,\omega_1,\omega_2,
\varphi_1,\varphi_2 \}$, where we have adopted a collective notation $\chi_k$, $k=1,2,\dots,6$ for the six
group parameters. We then integrate the associated volume form over the
parameters {\it not} in the maximal torus, resulting in a two-form $p(\theta_1,\theta_2)d\theta_1 d\theta_2$.
Normalizing $p(\theta_1,\theta_2)$ then gives the distribution $P(\theta_1,\theta_2)$.

We obtain the Haar measure by first computing the left-invariant\footnote{Of course, using the right-invariant
one-forms $R_k= \frac{\partial g}{\partial \chi_k} g^{-1}$ instead would lead to the same result.}
one-forms $L_k=g^{-1}\frac{\partial g}{\partial \chi_k}$,
also known as Maurer-Cartan forms, which take values in the Lie algebra of $SO(4)$. 
A left- and right-invariant volume form on the group manifold is then given by
\beq
\label{volumeform}
\sqrt{\det {\cal G}_{kl}}\, d^6\chi,
\eeq
which involves the ``metric"\footnote{$\cal G$ is symmetric and bilinear with non-negative eigenvalues, but
has degeneracies at some points $g\in SO(4)$.} ${\cal G}_{kl}$ constructed from the left-invariant one-forms
\beq
{\cal G}_{kl} =-\, {\rm Tr}(L_k L_l).
\eeq
Explicitly, one finds
\bea
\hspace{-0.5cm}
\det {\cal G}_{kl}&\!\! =\!\! &2^8\, (\cos\theta_1 -\cos\theta_2 )^4 (\cos (2\omega_1)-\cos(2\omega_2))^2\nonumber\\
&\! =\! & 2^{14}\,  \sin^4 \bigg( \frac{\theta_1 +\theta_2}{2}\bigg) \sin^4  \bigg(\frac{\theta_1 -\theta_2}{2}\bigg) 
\sin^2(\omega_1+\omega_2)\, \sin^2(\omega_1-\omega_2).
\eea
Since the determinant factorizes into a part depending on the torus variables $\theta_i$ and a rest, the same
holds for the volume form (\ref{volumeform}) and we can simply read off $p(\theta_1,\theta_2)$ up to
a proportionality constant. After normalization one finds the searched-for distribution of the angles $\theta_1$
and $\theta_2$,
\beq\label{1.7}
P(\th_1,\th_2) = \frac{1}{\pi^2} \,
 \sin^2 \bigg( \frac{\theta_1 +\theta_2}{2}\bigg) \sin^2  \bigg(\frac{\theta_1 -\theta_2}{2}\bigg).
\eeq
Below we will report on the measurement of the distribution $P(\th_1,\th_2)$ by
Monte Carlo simulations in CDT, and will find them to be in perfect agreement with formula \rf{1.7}.

\section{Implementation}
\label{impl}

We will now show how Wilson loops can be computed in the set-up of (causal) dynamical triangulations.
Despite the fact that we are in four dimensions, the process is entirely straightforward and easy to implement.
As explained earlier, we work with CDT building blocks that after the Wick rotation are equilateral,
and therefore are all identical with respect to their geometric properties.

Of the many possible coordinate systems for an individual four-simplex $s$ of this type, we choose an 
orthonormal frame whose origin coincides with the barycentre of $s$, which is equidistant to the five vertices of the 
simplex. For definiteness, we fix a scale such that the vertices all have geodesic distance 1 to the barycentre. This is
a coordinate-independent statement, which implies that the edge length of the simplex is $\ell=\sqrt{5/2}$.
After assigning labels $P_i$, $i=1,\dots,5$ to the five vertices of $s$, we fix the coordinate system uniquely by
choosing coordinates for the point $P_i$. Representing the four-tuple of coordinates of the $i^{\rm th}$ point
by a column vector $\vecx_i=\vec x (P_i)$, the explicit choice is  
\beq\label{2.1}
\begin{pmatrix}
\vecx_1,  \vecx_2,  \vecx_3,  \vecx_4,  \vecx_5
\end{pmatrix} = 
\begin{pmatrix}
0 & 0 & 0  & \gam & -\gam \\
0 & 0 &  \bt & -\frac{\bt}{2} & -\frac{\bt}{2}\\
0 &\al & -\frac{\al}{3} & -\frac{\al}{3} & -\frac{\al}{3}\\
1 & -\oq & -\oq & -\oq & -\oq
\end{pmatrix},
\eeq
with 
\beq\label{2.2}
\al = \frac{\sqrt{15}}{4}, ~~~~\bt= \sqrt{\frac{5}{6}},~~~~
\gam = \sqrt{\frac{5}{8}} .
\eeq
One easily verifies that the coordinate vectors satisfy
\beq\label{2.3}
\vecx_i^2 =1,~~\sum_i \vecx_i = \vec{0}, ~~\vecx_i \cdot \vecx_j = -\oq,
~i\neq j ,
\eeq
with respect to the usual scalar product of Cartesian coordinates on Euclidean space.\footnote{This
coordinate construction generalizes to $d$ dimensions, where the $d+1$ vertex coordinate vectors of an
equilateral $d$-simplex are required to satisfy the conditions (\ref{2.3}), with the third relation substituted by
$\vecx_i\cdot \vecx_j = -\frac{1}{d}$.} Our standard choice of coordinates on a simplex $s$ will be defined
through relations (\ref{2.1}), (\ref{2.2}) and (\ref{2.3}); given another simplex $s'$ with vertices $P'_i$, 
$i\! =\! 1,\dots, 5$, and coordinates $ x' $, its vertex coordinates will therefore be the same,
\beq
\label{idcoord}
\vec{x}'(P'_i) = \vec{x}(P_i).
\eeq
Since the standard coordinates 
depend on a specific labelling $\{ P_i\}$ of the vertices, we are left
with a residual, discrete coordinate freedom, associated with relabelling those vertices. The latter is given in
terms of the permutation group $S_5$ of five elements. Given such a permutation $\pi:\ i\mapsto \pi (i)$, there is
an associated permutation of vertices $P_i\mapsto P_{\pi (i)}$, which in turn corresponds to a linear
transformation ${\cal P}_\pi$ of the coordinate system. Adhering to the column vector notation
introduced earlier, ${\cal P}_\pi$ is a $(4\times 4)$-matrix given by
\beq
\label{permlin1}
\begin{pmatrix}\vecx_{\pi_1,}  \vecx_{\pi_2},  \vecx_{\pi_3},  \vecx_{\pi_4}\end{pmatrix}
 = \cP_\pi \cdot \begin{pmatrix}
\vecx_1,  \vecx_2,  \vecx_3,  \vecx_4
\end{pmatrix}.
\eeq
Since the four vectors $\vecx_i$, $i=1,\dots, 4$ are linearly independent, we can solve this equation to obtain
\beq
\label{permlin2}
~ \cP_\pi = \begin{pmatrix}\vecx_{\pi_1,}  \vecx_{\pi_2},  \vecx_{\pi_3},  \vecx_{\pi_4}\end{pmatrix}\cdot 
\begin{pmatrix}
\vecx_1,  \vecx_2,  \vecx_3,  \vecx_4
\end{pmatrix}^{-1}.
\eeq 
Since all scalar products are preserved by virtue of the relations (\ref{2.3}), this transformation is necessarily
orthogonal, ${\cal P}_\pi \in O(4)$.\footnote{Recall that $O(n)$ has two connected components, $SO(n)$ being the 
component connected to the identity. The determinant of any 
member of the other component is $-1$.} In other words, we have obtained a representation of the permutation
group $S_5$ in terms of orthogonal matrices in four dimensions.
If the permutation is even, we have $\det {\cal P}_\pi\! =\! 1$ and 
${\cal P}_\pi \in SO(4)$, if the permutation is odd, we have $\det {\cal P}_\pi \! =\! -1$ and
${\cal P}_\pi \notin SO(4)$.

Having made a coordinate choice for a given four-simplex and vertex labelling, we will now address
the explicit construction and computation of the holonomy $R_L$ associated with a 
closed loop $L$ passing through a sequence of
four-simplices, along an oriented, piecewise straight path through adjacent simplex (bary)centres, as described in 
Sec.\ \ref{holodt} above. Note that a path that passes through a simplex $s$ distinguishes a triangle in
$s$, namely, the triangle shared by the two faces (tetrahedra) through which the path enters and leaves $s$.
One may view this triangle as the hinge around which the path ``bends''. If the path follows a complete set of 
consecutive four-simplices which share a single interior triangle $\tau$ of a triangulation, 
we are back to the situation of a minimal
loop, whose associated holonomy -- up to conjugation -- is a rotation matrix of the form \rf{1.6}.
 
The task at hand is to construct the rotation matrices $R(s_{i+1},s_i)$ in the expression 
for the holonomy (\ref{1.1}) of a loop passing through the 
simplices $s_1,s_2,\ldots,s_n,s_1$. Having fixed a standard coordinate system for a given labelling of the
vertices, the remaining gauge freedom we have to compute $R_L$ 
is how to pick the vertex labels for the members of the set $\{ s_i \}$.
For illustration, we will consider two different ways of computing the rotation matrix $R(s_{i+1},s_i)$ 
associated with two adjacent simplices $s_i$ and $s_{i+1}$ in the sequence. In the first one, the
vertex labelling of simplex $s_{i+1}$ is related to that of the previous simplex $s_i$ along the loop $L$,
and in the second one, the labellings of $s_i$ and $s_{i+1}$ are picked independently and arbitrarily 
beforehand. It is the latter we will use in the simulations later on.

\subsection{Choosing vertex labels (version 1)}

\begin{figure}[t]
\centering
\scalebox{0.5}{\includegraphics{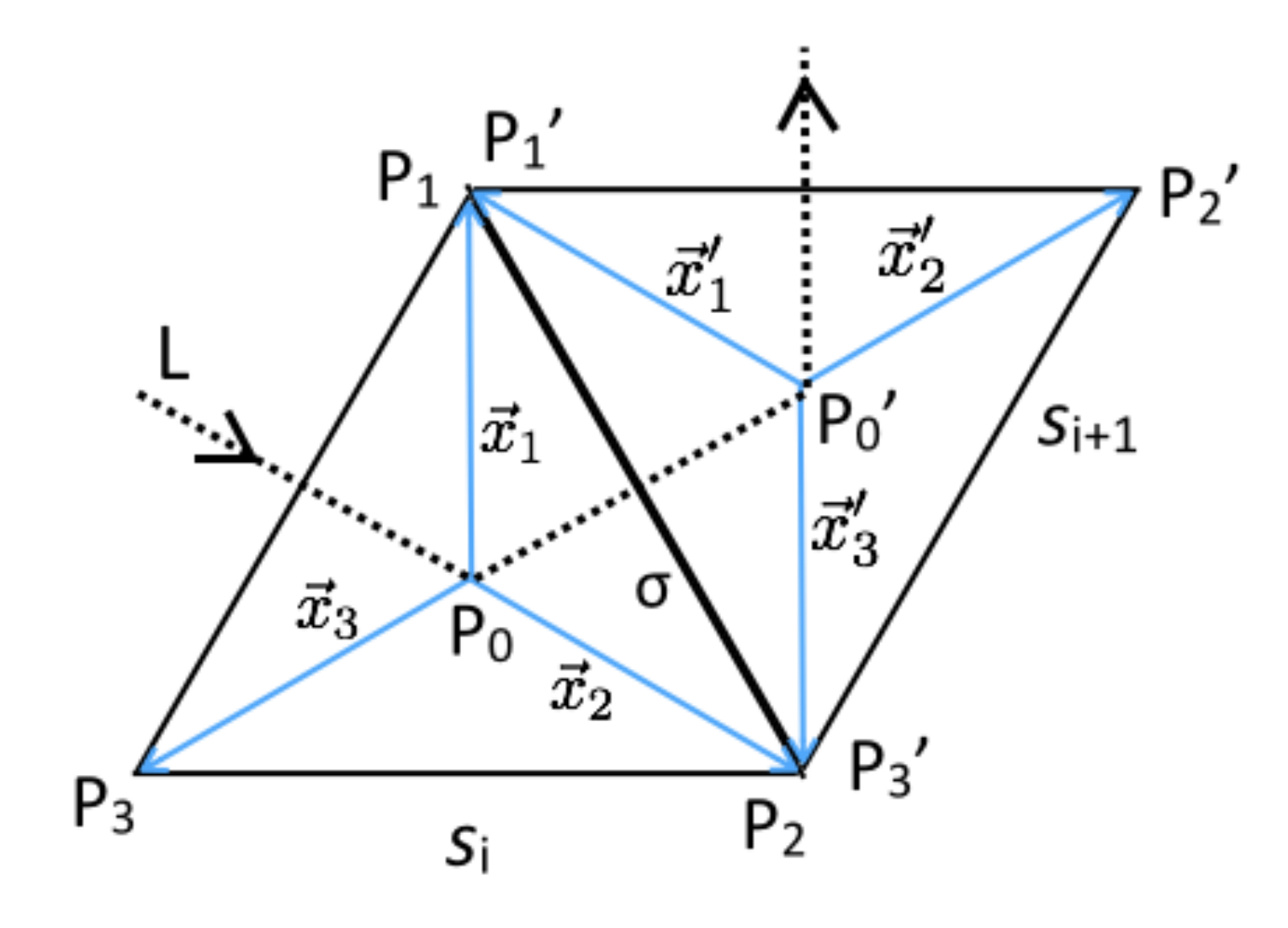}}
\caption[phase]{Two-dimensional analogue of the coordinate assignments for consecutive
simplices $s_i$, $s_{i+1}$ (here represented by triangles) traversed by part of an oriented loop $L$.
The vertex labels are $\{P_1,P_2,P_3\}$ and $\{P'_1,P'_2,P'_3\}$, as indicated. 
The triangles share a common face $\sigma$ (the edge between $P_1$ and $P_2$),
and the common hinge around which the rotation takes place is the vertex $P_1\! =\! P_1'$.
The coordinate system $\{ x^\mu\}$ based at $P_0$ 
is defined in terms of three coordinate unit vectors $\vec x_i$ for the vertices $P_i$, $i=1,2,3$, and
the coordinate system $\{ x'^\mu \}$ is defined analogously in terms of primed quantities. 
The two-dimensional counterparts
of relations (\ref{2.4}) and (\ref{zdet}) are given by 
$\vec x'_1 =\vec x_1$, $\vec x'_2 =\vec x_3$ and $\vec z=-\vec x_2$. 
}     
\label{coordinate}
\end{figure}

Let the vertices of $s_i$ be labeled by $\{1,2,3,4,5\}$, with standard coordinates 
\rf{2.1} assigned to them. Assume that the boundary tetrahedron shared by $s_i$ and $s_{i+1}$
has vertex labels $\{1,2,3,4\}$, and that the distinguished triangle $\tau$
around which the holonomy ``rotates'' has vertex labels $\{1,2,3\}$. 
Now choose the vertex labels in simplex $s_{i+1}$ such that for the vertices shared
by both four-simplices we have
\beq
P_i=P_i',\;\;\; i=1,2,3,\;\;\;\; P_4=P_5',
\eeq
where the primed vertices refer to simplex $s_{i+1}$. Using our standard coordinates, 
this fixes uniquely a coordinate system $\{ x'^\mu \}$ in $s_{i+1}$, and the vertex coordinates of
the two simplices are related by
\beq\label{2.4}
\vecx'(P_i) = \vecx(P_i),~~i=1,2,3,~~~~\vecx'(P_4) = \vecx(P_5).
\eeq
(Fig.\ \ref{coordinate} illustrates the analogous situation in two dimensions.)
To compute the matrix $R(s_{i+1},s_i)$, we make use of the observation 
from Sec.\ \ref{holodt} that the coordinate system $\{ x^\mu \}$ thus defined in $s_i$ 
naturally extends to the neighbouring simplex $s_{i+1}$, and vice versa for the
coordinates $\{ x'^\mu \}$. The relation between the $x$- and the $x'$-coordinates of the same
point $P$ involves the rotation matrix $R$ and a translation (c.f. Fig.\ \ref{coordinate})
according to
\beq\label{2.6}
\vecx'(P) = R(s_{i+1},s_i) \,\vecx(P) + \vec{z},
\eeq
where $\vec z$ is the difference vector between the barycentres $P_0$ and $P_0'$.
To determine the matrix $R(s_{i+1},s_i)$ it suffices to know the coordinates of
four vertices in both coordinate systems, as well as the vector $\vec z$. 
We can determine $\vec z$ by setting $P\! =\! P_0$ in eq.\ (\ref{2.6}), and using
the fact that $\vec x(P_0)$ is the zero vector, because by definition the 
barycentre $P_0$ is the origin of the coordinate system $\{ x^\mu\}$.
Furthermore, one can work out by elementary trigonometry that $P_0$ has the
$x'$-coordinates
\beq
\label{locbary}
\vecx'(P_0) = -\frac{1}{2}\, \vecx'(P_4').
\eeq
We thus obtain from eq.\ (\ref{2.6})
\beq
\label{zdet}
\vec z= 
-\frac{1}{2}\, \vecx'(P_4') =-\frac{1}{2}\, \vec x (P_4),
\eeq
where the last equality holds because of the identities (\ref{idcoord}).\footnote{The analogous
relation in $d$ dimensions is given by 
$\vec z=-\frac{2}{d} \vec x (P_d)$.}
Collecting all the information, using relations (\ref{2.4}) and introducing the shorthand notation
$\vecx_{i}:=\vecx(P_{i}),~~\vecx'_{i}:=\vecx'(P_{i})$, we obtain a complete set of equations for 
the rotation matrix $R(s_{i+1},s_i)$, namely,
\beq
\label{matr}
\vec x_i+\frac{1}{2}\, \vec x_4=R(s_{i+1},s_i)\, \vec x_i,\;\; i=1,2,3,\;\;\;\;\;\;
\vec x_5+\frac{1}{2}\, \vec x_4=R(s_{i+1},s_i)\, \vec x_4.
\eeq 
The above construction can be generalized immediately to an arbitrary permutation 
$\{n_1,n_2,n_3,n_4,n_5\}$ of the vertex labels $\{1,2,3,4,5\}$ for the simplex $s_i$,
with $\{n_1,n_2,n_3,n_4\}$ denoting the vertices of the tetrahedron shared with simplex
$s_{i+1}$ and $\{n_1,n_2,n_3\}$ the vertex labels of the distinguished triangle. 
The relations (\ref{2.4}), (\ref{locbary}) and (\ref{zdet}) become  
\beq\label{2.4a}
\vecx'_{n_i} = \vecx_{n_i},~~i=1,2,3,~~~~~\vecx'_{n_4} = \vecx_{n_5},
~~~~~
\vecx'_0 = -\oh\, \vecx_{n_4}.
\eeq 
Using a column vector notation for the coordinate four-tuples, the defining relation for the rotation 
matrix $R$ can be written as the matrix equation
\beq\label{2.7}
\begin{pmatrix}
\vecx_{n_1}\! +\!\oh \vecx_{n_4} ,\vecx_{n_2}\! +\! \oh \vecx_{n_4} ,
\vecx_{n_3}\! +\! \oh \vecx_{n_4} , \vecx_{n_5}\! +\! \oh \vecx_{n_4}  
\end{pmatrix} = R(s_{i+1},s_i)\! \cdot \! 
\begin{pmatrix}
\vecx_{n_1}  , \vecx_{n_2} , \vecx_3 , \vecx_{n_4}
\end{pmatrix},
\eeq
generalizing equations (\ref{matr}).
One can check that the solution $R$ to eq.\ (\ref{2.7}) does not depend on the permutation
of the three triangle indices and that $R\in SO(4)$. More specifically, up to conjugation 
associated with a permutation of the five vertex labels of $s_i$, $R$ is equivalent to the matrix
\beq
\label{specialmat}
\begin{pmatrix} 
1 & 0 & 0&0 \\
0 & 1 & 0&0 \\
0 &0& \frac{1}{4} & \frac{\sqrt{15}}{4}  \\
0 & 0 & - \frac{\sqrt{15}}{4} & \frac{1}{4} 
\end{pmatrix},
\eeq
which according to our discussion in Sec.\ \ref{holodt} is an example of a simple rotation,
in the present case representing a rotation by an angle $\theta\! =\!\arccos \frac{1}{4}$ in the plane 
perpendicular to the triangular hinge.

Starting at the initial simplex $s_1$, we can apply the procedure just described iteratively 
to choose vertex labels and associated coordinate systems for all other simplices traversed
by the loop $L$. However, we are {\it not} guaranteed to arrive back at $s_1$ with the same
vertex labelling we started out with, but instead will end up with some label set 
$\{ \al_1,\al_2,\al_3,\al_4,\al_5 \}$. The last rotation matrix in the sequence, $R(s_1,s_n)$,
therefore has to be followed by the rotation matrix ${\cal P}_\pi$ implementing the 
permutation $\pi$ which brings the vertices back to the labelling chosen in $s_1$ at the outset,
that is, for which
\beq
\pi (\al_i)\! =\! i,\;\; i=1,2,\dots, 5,\;\;\; \Longrightarrow \;\;\;
\cP_\pi \! \cdot\!\begin{pmatrix}\vecx_{\al_1} , \vecx_{\al_2}, \vecx_{\al_3} , \vecx_{\al_4}
\end{pmatrix}=
\begin{pmatrix}
\vecx_1 , \vecx_2, \vecx_3, \vecx_4
\end{pmatrix}.
\eeq
For the given choice of vertex labels, we have finally arrived at the concrete expression for the holonomy matrix $R_L$;
it is given by
\beq\label{2.9}
R_L = \cP_\pi\, R(s_1,s_n) \cdots R(s_2,s_1) .
\eeq

\subsection{Choosing vertex labels (version 2)}

Any choice of coordinate systems for the simplices along a loop $L$ will affect the holonomy matrix
$R_L$ at most by conjugation and the invariants (\ref{trace}) and (\ref{tracesq}) not at all. 
The procedure outlined in the previous subsection was natural in the sense that 
aligning the coordinate systems of successive four-simplices led to a simple
geo\-metric interpretation of the matrices $R$ as being associated with the rotations around 
the triangles singled out by the loop $L$. However, from a computer point of view 
it is slightly inconvenient to have to identify these triangular hinges and to label the vertices of
the four-simplices anew for each new path $L$, and finally to compute the permutation matrix
${\cal P}_\pi$. 

It turns out to be computationally advantageous to define local coordinate systems which 
make use of the fact that as part of the Monte Carlo set-up each four-simplex already comes 
with a labelling of its vertices in terms of a permutation of the indices $\{1,2,3,4,5\}$. We have to
generalize our considerations of the previous section only slightly to obtain the holonomy matrix
for this case. 

Consider again the transition from simplex $s_i$ to simplex $s_{i+1}$. The general situation is
that the four vertices spanning the common tetrahedral face between the two four-simplices
have labels $\{n_1,n_2,n_3,n_4 \}$ in $s_i$ and labels $\{k_1,k_2,k_3,k_4 \}$ in $s_{i+1}$. 
Let $n_5$ and $k_5$ be the remaining labels of the fifth vertex in $s_i$ and $s_{i+1}$ respectively.
The transformation matrix $R(s_{i+1},s_i)$ we are looking for satisfies 
\beq\label{2.10}
\begin{pmatrix}
\vecx_{k_1}\! +\! \oh \vecx_{k_5}, \vecx_{k_2}\! +\! \oh \vecx_{k_5} ,
\vecx_{k_3} \!+\! \oh \vecx_{k_5}, \vecx_{k_4}\! +\!\oh \vecx_{k_5}  
\end{pmatrix} \! =\! R(s_{i+1},s_i) \! \cdot \!
\begin{pmatrix}
\vecx_{n_1}, \vecx_{n_2}, \vecx_{n_3}, \vecx_{n_4}
\end{pmatrix},
\eeq
in analogy with eq.\ \rf{2.7}. From this equation we can read off that the solution 
\beq\label{2.11} 
R(s_{i+1},s_i)\! =\!   
\begin{pmatrix}
\vecx_{k_1} \! +\! \oh \vecx_{k_5}, \vecx_{k_2}\! +\! \oh \vecx_{k_5},
\vecx_{k_3} \!+\! \oh \vecx_{k_5}, \vecx_{k_4}\! +\! \oh \vecx_{k_5} 
\end{pmatrix}\! \cdot \! \begin{pmatrix}
\vecx_{n_1}, \vecx_{n_2}, \vecx_{n_3}, \vecx_{n_4}
\end{pmatrix}^{-1} 
\eeq
is invariant under simultaneous permutations of the label sets $\{n_1,n_2,n_3,n_4 \}$ and 
$\{k_1,k_2,k_3,k_4 \}$. 

Note that in the computer program no attention is paid to the relative orientation of the four-simplices when 
labelling their vertices with a permutation of $\{1,2,3,4,5\}$. This implies that a matrix $R(s_{i+1},s_i)$ 
computed from (\ref{2.11}) can have determinant $-1$ rather than 1, i.e.\
belong to $O(4)$ rather than $SO(4)$, which according to Sec.\ \ref{holodt} is the generic case anyway.
However, for a closed curve $L$ there will always be an even number of $O(4)$-matrices 
with determinant $-1$ and the holonomy matrix $R_L$ will therefore always be a $SO(4)$-matrix.
In terms of the new $R$-matrices determined from \rf{2.10} we can write the  
holonomy as
\beq\label{2.12}
R_L' =  R(s_1,s_n) \cdots R(s_2,s_1).
\eeq
This is the analogue of \rf{2.9}, but without the need for a
permutation matrix $\cP_\pi$, since we will always return to the same labelling of the vertices of 
the simplex $s_1$ where the loop $L$ starts and ends. We will use the prescription leading to
expression (\ref{2.12}) in the measurement of holonomies described in Sec.\ \ref{measure} below.

By construction only a finite number of $R$-matrices can occur. They can be computed, stored
and looked up for given pairs of vertex label sets, without having to perform the matrix inversion 
and multiplication of formula (\ref{2.11}) each time. In addition, we can make maximal use of the
permutation invariance mentioned above. For example, we only need the five matrices
\bea 
&&
\hspace{-0.8cm}
\begin{pmatrix}
\vecx_{1}, \vecx_{2}, \vecx_3, \vecx_{4}
\end{pmatrix}^{-1} 
= 
\begin{pmatrix}
\al_1 &\bt_1 &\gam_1& 1 \\
\al_1&\bt_1 & 4\gam_1 & 0\\
\al_1 & 3\bt_1& 0&0\\
2\al_1&0&0&0
\end{pmatrix}
,\;
\begin{pmatrix}
-\al_1 &\bt_1 &\gam_1& 1 \\
-\al_1&\bt_1 & 4\gam_1 & 0\\
-\al_1 & 3\bt_1& 0&0\\
-2\al_1& 0 & 0 & 0
\end{pmatrix},
\\
\vspace{0.3cm}\nonumber
\\
&&
\hspace{-0.2cm} 
\begin{pmatrix}
0 & -2\bt_1 &\gam_1& 1 \\
0 &-2\bt_1 & 4\gam_1 & 0\\
\al_1 & -3\bt_1 & 0 & 0\\
-\al_1 &-3\bt_1 &0 &0
\end{pmatrix}
,\;
\begin{pmatrix}
0 &0  &-3\gam_1 & 1 \\
0&2\bt_1 & -4\gam_1 & 0\\
\al_1 & -\bt_1& -4\gam_1 &0\\
-\al_1& -\bt_1& -4\gam_1& 0
\end{pmatrix}
,\; 
\begin{pmatrix}
0 & 0 &3\gam_1& -1 \\
0& 2\bt_1 & -\gam_1 & -1\\
\al_1 & -\bt_1& -\gam_1 & -1\\
-\al_1 & -\bt_1 & -\gam_1 & -1
\end{pmatrix}
\nonumber
\label{2.17}
\eea
with 
\beq\label{2.18}   
\al_1 = \sqrt{\frac{2}{5}},~~~\bt_1= \sqrt{\frac{2}{15}},~~~
\gam_1=  \sqrt{\frac{1}{15}},
\eeq
when determining $R(s_{i+1},s_i)$,
because we can always find a simultaneous permutation of $\{n_1,n_2,n_3,n_4 \}$ and 
$\{k_1,k_2,k_3,k_4 \}$ to make the second matrix on the right-hand side of eq.\ (\ref{2.11})
match one of these five matrices.

\section{Wilson lines in CDT}
\label{wilsonlines}

Having given a concrete prescription for the computation of the transition matrices $R(s_{i+1},s_i)$ 
and how to obtain from them the holonomy $R_L$ associated with a given lattice loop $L$, let us now
turn to what we can learn about the behaviour of the holonomies and their potential use in probing 
``quantum geometry", that is, the geometric properties of the dynamically generated ground state of the 
nonperturbative path integral.

One possibility would be to study the holonomy group of a given, fixed piecewise flat 
geometry, taken to be a typical member of the gravitational 
path integral ensemble generated by the Monte Carlo simulations.
We know that these geometries are non-differentiable and highly singular (similar to the 
configurations -- the ``paths" -- of an ordinary non-relativistic quantum-mechanical 
path integral), and one could compare
the holonomy groups of different, fixed such geometries with those of smooth Riemannian manifolds. 

We will be interested here in the physically more interesting case of 
using holonomies to construct observables in 
the fully dynamical, nonperturbative quantum theory. By ``observables" we mean in this context 
coordinate-invariant quantities, which are operationally well-defined on the quantum-fluctuating
ensemble of geometries. We do not require them to be related explicitly to any truly observable 
phenomenological effects (other than perhaps in some semiclassical limit), which 
would be a tall order in any theory of quantum gravity. An example of an
observable in this looser sense is the (expectation value of the)
spectral dimension of quantum spacetime, a quantity which has
been measured explicitly in CDT quantum gravity \cite{spectral}, and also studied in other 
formulations \cite{spectralother}. 

Coming up with physically interesting observables in the sense just described is still a formidable 
challenge in background-free, nonperturbative quantum gravity. To illustrate the point, 
consider some two-point function $G_2(x,y)$ in standard quantum field theory on a {\it fixed} background.
Its na\"ive analogue on a nonperturbatively fluctuating ensemble of geometries is not a well-defined
observable, because there is no coordinate-invariant way to refer to {\it the same} two points $x$ and $y$
throughout the ensemble. One workaround 
is to specify the geodesic distance of the
two points to be $r$ and integrate over all possible positions of $x$ and $y$ subject to this constraint.
Averaging this quantity in the path integral over the geometric ensemble then leads to a well-defined 
two-point function 
\beq
{\cal G}_2(r)\! =\!\!\int\! {\cal D}[g_{\mu\nu}]\mbox{e}^{-S[g_{\mu\nu}]}\!\! \int\!\! dx\, dy\sqrt{g(x)g(y)} 
G_2(x,y)\delta(r\! -\! d_{g_{\mu\nu}}\! (x,y))
\label{diffeoinvprop}
\eeq
(see \cite{ambagn} for a concrete implementation in two dimensions). 
The dependence of the diffeomorphism-invariant propagator ${\cal G}_2$ on the coordinate-invariant 
geodesic distance $r$ captures nontrivial physical information. 

There are similar issues when trying to construct an observable that depends not just on two points, but
on an entire closed curve in spacetime. An obvious generalization of the prescription leading to a
well-defined two-point function would be to consider the trace invariants (\ref{trace}), (\ref{tracesq}) 
for a subclass of
loops sharing certain invariant geometric features in terms of their length and shape, and then to 
integrate over all possible locations of such loops. 

In the present work, we will pursue a similar strategy to construct and measure a particular class of 
well-defined 
Wilson loop observables, but instead of referring to intrinsic geometric properties of the underlying
paths, we will introduce a dynamical point particle, couple it to the quantum geometry, and compute
the holonomy along its world line. In other words, we will consider an interacting system of matter and
geometry, where a massive point particle is coupled to pure quantum gravity, given in terms of
the usual CDT ensemble of fluctuating geometries. The (Euclidean) action of a point
particle of mass $m$ with spacetime trajectory $\gamma$ in a Riemannian geometry with metric 
$g_{\mu\nu}$ is given by the mass times the proper length  of $\gamma$,
\beq\label{3.1}
S_{\gamma}^{\rm \, p.p.} = m\! \int_\gamma dl,~~~~dl = \sqrt{g_{\mu\nu}(\gamma (s)) 
\frac{d\gamma^\mu}{ds} \frac{d\gamma^\nu}{ds}}\; ds.
\eeq
Concretely, we will add an appropriate simplicial version of the point particle action
(\ref{3.1}) to the Einstein-Hilbert action and update the combined, interacting system
using Monte Carlo simulations, as described in Sec.\ \ref{measure} below.

As usual in CDT, we consider spacetimes 
of topology $S^3\times S^1$, where -- for convenience -- time has been compactified
to a circle.
The particle world lines whose holonomies we will measure are oriented in 
positive time direction and wind
around the periodic time direction exactly once. Accordingly, our loops all have
macroscopic length, with a minimum that depends on the time extension of the spacetime.
In line with standard terminology from gauge theory
we will refer to these holonomies and their associated trace invariants (the context should
make clear which is meant) as Wilson lines. 

To understand the geometry of these Wilson lines, we need to recap briefly some
aspects of the triangulated spacetimes in the configuration space of the
path integral. More complete descriptions can be found elsewhere, 
see \cite{physrep} and references therein.
In standard CDT, each spacetime $\cal T$ has a proper-time slicing with integer label $t$, 
and is assembled from four-simplices in a layered fashion\footnote{see \cite{jordanloll} for a
generalization of CDT geometries, without strict time slicing, but maintaining causality}, where one layer of
thickness $\Delta t\! =\! 1$ is a piecewise flat piece of spacetime of topology $S^3\times I$,
all of whose vertices are contained in either of its spatial boundary submanifolds at times 
$t$ or $t+1$. These submanifolds are arbitrary triangulations in terms of equilateral
tetrahedra, and all have the topology of a three-sphere.
An entire four-geometry of proper-time extension $T$ is obtained by
gluing together $T$ subsequent layers along matching three-geometries,
and finally identifying the final boundary of the last layer with the initial boundary of
the first layer.

\begin{figure}[t]
\centerline{\scalebox{0.6}{\rotatebox{0}{\includegraphics{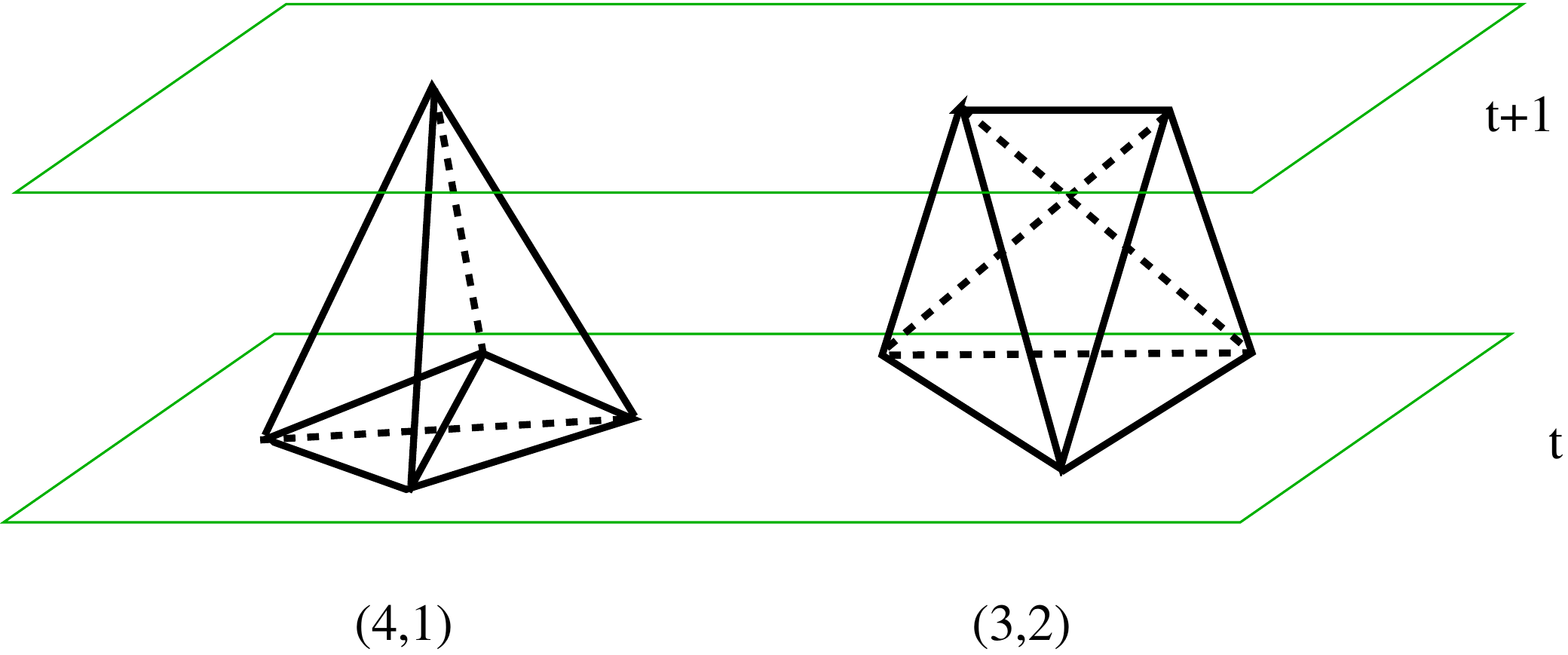}}}}
\caption[phased]{The two types of elementary simplicial building blocks used in CDT quantum gravity,
and how they are positioned with respect to two adjacent slices of constant proper time. Together with the
(4,1)- and the (3,2)-simplex shown, also their time-reversed versions, the (1,4)- and the (2,3)-simplex, appear.}
\label{foursimpl}
\end{figure}

Before the Wick
rotation, we distinguish between space- and timelike edges. The former are always contained
in a three-dimensional spatial slice of fixed proper time, whereas the latter interpolate 
between two adjacent spatial slices of fixed proper times $t$ and $t+1$. 
There are (up to time reflection) two types of elementary building blocks, the (4,1)-simplex
and the (3,2)-simplex, see Fig.\ \ref{foursimpl}. 
The notation indicates how they are positioned in a given layer with respect
to the slices of constant integer time: a four-simplex of type $(i,j)$ has $i$ vertices with time label $t$ and
$j$ vertices with time label $t+1$. Building blocks of different types have different numbers of 
time- and spacelike edges. Although after the Wick rotation all edges are spacelike and -- with
the convention adopted in this paper -- of equal length $\ell$, one can still distinguish the
different building blocks because of how they are positioned inside the layered structure, 
since the Wick rotation leaves the connectivity of the triangulation intact. 

\section{Measurements}
\label{measure}

As usual in CDT simulations, we keep the number $T$ of proper-time steps fixed, as well as the
total four-volume (defined as the total number $N_4$ of four-simplices), which turns out to be
technically convenient. For the measurements reported below, we worked with $T=80$ and
$N_4=20.000$, at the point $(\kappa_0,\Delta)=(2.2,0.6)$ in the space of coupling constants, 
which lies inside the ``de Sitter phase" of CDT quantum gravity \cite{agjl} ($\kappa_0$ is the bare
inverse Newton constant, and $\Delta$ denotes the so-called asymmetry parameter).
The action we use is the Regge form of the Einstein-Hilbert action \cite{regge},  
applicable to piecewise linear geometries. On causal dynamical triangulations with identical
equilateral building blocks it assumes a particularly simple form (see \cite{physrep} for details). 

The Monte Carlo computer simulations proceed from an initial configuration put in by hand,
after which we apply a standard Metropolis algorithm to update the geometry,
using as action the Regge-Einstein-Hilbert action, and allowing only those
Monte Carlo moves (changes in the geometry) that are compatible with the foliated structure
of the spacetimes. The simulations generate a sequence of four-dimensional geometries -- 
spacetime histories represented by triangulations -- which after sufficiently many updates 
will be independent of the chosen starting configuration. 
Computing the expectation value of a given observable amounts to measuring the observable many times for 
statistically independent geometries generated by the Monte Carlo simulation, and calculating
its average inside the path integral.

We describe next how to add a Wilson line to the Monte Carlo simulations of pure gravity.
In the selected initial configuration, we put in by hand a closed path that winds once around
the periodic time direction. Without loss of generality, we follow the choice made in 
Sec.\ \ref{holodt} and consider only piecewise straight paths connecting the centres of
neighbouring four-simplices. It is important to understand that not every type of four-simplex
can be a neighbour of every other type of four-simplex, in the sense of having a three-simplex
in common. The reason for this is that {\it before} the Wick rotation the three-simplices also
come in different types, depending on the time- and spacelike character of their edges, and
that gluings of two four-simplices along two tetrahedral faces are only possible if
the metric properties of the tetrahedra match exactly. For this reason, a (4,1)-simplex cannot
be a neighbour of a (2,3)-simplex, say.

For a path moving forward in time this implies that only
particular sequences of the simplex types can occur along it, namely, those of the form
\beq
\label{seq}
\dots \underbrace{(4,1),\dots ,(4,1)}_{\text{$m_1$ times, $m_1\geq 1$}},
\underbrace{(3,2),\dots ,(3,2)}_{\text{$m_2$ times, $m_2\geq 1$}},
\underbrace{(2,3),\dots ,(2,3)}_{\text{$m_3$ times, $m_3\geq 1$}},
\underbrace{(1,4),\dots ,(1,4)}_{\text{$m_4$ times, $m_4\geq 1$}},
\underbrace{(4,1),\dots,}_{\text{$m_5$ times}} 
\eeq
which should be continued cyclically and read as: a (4,1)-simplex can only be followed by another (4,1)-simplex or
by a (3,2)-simplex, a (3,2)-simplex can only be followed by another (3,2)-simplex or
by a (2,3)-simplex, and so forth. It follows that such a path needs to go through at least four four-simplices
to pass from one layer to the next, and to arrive at the same type of four-simplex it started from.
For example, consider a path starting at the barycentre of a (4,1)-simplex in the layer between
times $t$ and $t+1$. It has to pass through at least three other four-simplices in the same layer 
before arriving at a (4,1)-simplex in the layer between times $t+1$ and $t+2$, namely, one
(3,2)-simplex, one (2,3)-simplex and one (1,4)-simplex. In other words, at least four steps are
necessary to advance by one time unit $\Delta t=1$.\footnote{Because of this substructure, one can
make a further subdivision of time, with units of 1/4, which is sometimes useful. 
This was first introduced in \cite{volprofile}, see also \cite{physrep}.} 

In our set-up, an oriented path associated with a Wilson line is not allowed to
go back in time relative to the proper-time foliation and the sequence defined in (\ref{seq}), 
thereby enforcing some degree of ``causality".  
For example, it may move among the (3,2)-simplices of a given layer, but
not subsequently go back to a (4,1)-simplex of the same layer; it can only proceed to
a (2,3)-simplex, as specified by (\ref{seq}). Another restriction put on the paths is
that they are not allowed to self-intersect, mainly to prevent them from meandering for
very long times inside a given layer.

The Boltzmann weights of the combined configurations of four-geometry and particle path
now contain an additive contribution to the Einstein-Hilbert action, which is the lattice version of the
continuum action (\ref{3.1}) and simply given by
\beq\label{3.2} 
S_{L}^{\rm \; p.p.} = m_0 N_L,
\eeq
where $m_0$ is the bare particle mass and $N_L$ the number of four-simplices 
encountered by the loop $L$.
During the Monte Carlo simulations, the path evolves in computer time along with the geometry.
The usual local update moves are performed on the geometry. Whenever the particle world line
happens to pass through simplices affected by such a local rearrangement, it will be broken up
there. Part of the updating algorithm is then to determine all possible ways in which the two loose ends of the
path can be reconnected, and weighing them with the appropriate Boltzmann weights. 
Lastly, moves will be disallowed if they lead to a path which self-intersects or goes backward 
in time.\footnote{We will describe the Monte Carlo 
simulations of a point particle coupled to CDT quantum gravity
in detail elsewhere, where we also investigate systematically the effect of 
the particle on the quantum geometry and vice versa. In the present article we focus on the 
construction and measurement of the Wilson lines associated with the particle paths,
keeping the account of the technicalities of the computer simulations to a minimum.}

\begin{figure}[t]
\centerline{\scalebox{0.8}{\rotatebox{0}{\includegraphics{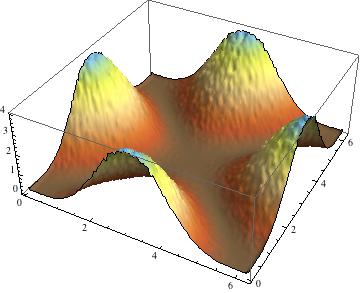}}}}
\caption[phased]{
Histogram of the invariant angles $(\th_1,\th_2)$ characterizing
a SO(4)-holonomy $R_L$, from Monte Carlo measurements.}
\label{fig1}
\end{figure}
The behaviour of the particle path $L$ depends on the parameter $m_0$ in 
the action $S_L^{\rm \; p.p.}$. For large $m_0$, long paths are
strongly suppressed, and the length will be $N_L\! =\! 4T\! =\! 360$, which minimizes the action \rf{3.2}. 
The constraints forbidding backtracking in time and self-intersections of $L$ are in this
case irrelevant, because such configurations occur only very rarely in the 
computer simulations. As $m_0$ becomes smaller, longer loops become less costly
and the length of $L$ will become longer and fluctuate more. As a consequence, the constraints
play a nontrivial role in limiting the length of the loop.    

Once the simulations are running and have thermalized, we sample configurations,
measure $R_L$ along the particle world line and compute the trace invariants $t_1$ and $t_2$
from (\ref{trace}) and (\ref{tracesq}).
This allows us to determine the angles $\th_1$ and $\th_2$ in the interval $[0,2\pi]$ up 
to an interchange $\th_1 \!\leftrightarrow \!\th_2$ and up to the reflections $\th_i \to 2\pi \! -\!\th_i$. 
Fig.\ \ref{fig1} shows the histogram of the measured values for the angles $\theta_i$. 
Its shape is independent of the mass $m_0$ in the point particle action \rf{3.2}, and by 
construction displays the above exchange and reflection symmetries. 
We have verified that after normalization the measured distribution is in 
perfect agreement with the theoretical 
distribution $P(\th_1,\th_2)$ of \rf{1.7}, which was derived under the assumption that the holonomy $R_L$ is 
uniformly distributed over the group manifold $SO(4)$.

\section{Discussion and outlook}
\label{discussion}

One could have wondered a priori whether the coordinate-free set-up of (Causal) Dynamical Triangulations
is suited to describing holonomies and Wilson loops. In this article, we have demonstrated conclusively that it is
straightforward to define and compute these quantities. More than that, we find it difficult to envisage
a framework, lattice-based or otherwise, that would make the computation of Wilson loops on four-dimensional 
curved manifolds even simpler. 
Our explicit construction involved a particular choice of Cartesian coordinate systems on the 
individual simplicial building blocks, but clearly many other choices are possible and would not affect the final result, 
which was formulated in terms of coordinate-independent quantities. 

One could also have worried a priori that the computation of holonomies in the (C)DT framework was affected
strongly by discretization effects, especially since the equilateral simplices have just a single interior angle 
$\alpha_{\rm int}\! =\! \arccos 1/4$ (the angle between two three-dimensional faces sharing a two-dimensional hinge),
despite the fact that this angle is irrational. In addition, we saw in Sec.\ \ref{impl} that only a finite number of
different transition matrices $R(s_{i+1}, s_i)$ occur in the holonomy computations. 
However, we have not observed any sign of such discretization artefacts. 
On the contrary, our main result is that for the class of Wilson lines considered, the holonomies appear to
cover SO(4) densely and uniformly. This also implies that the holonomy group of the quantum geometry generated in the
de Sitter phase of CDT quantum gravity is SO(4), which is the same as the holonomy group of a generic orientable
four-dimensional Riemannian manifold. Conversely, we have not found any tendency of our macroscopic Wilson
lines to cluster around the identity of the group SO(4).

We have shown that the CDT framework is well suited for investigating Wilson loops in nonperturbative
quantum gravity. Our Wilson lines are well-defined observables, but we have at this stage no direct
physical interpretation to relate them to specific classical or quantum properties of the underlying
quantum spacetime. {\it The} challenge for any theory of quantum gravity is to come up with observables
which do this. As already mentioned in the Introduction, quantities involving holonomies in one way or 
other are natural candidates for encoding information about the curvature of (quantum) spacetime. 
Our analysis of the holonomy of minimal loops around a single triangular hinge shows that
the classical relation between the holonomy of an infinitesimal planar loop and the local curvature -- here
in the form of a deficit angle \`a la Regge --
continues to hold on the piecewise flat geometries of (Causal) Dynamical Triangulations.

From the point of view of the regularized lattice formulation, these minimal loops are not particularly
interesting, since they merely probe geometry at the cutoff scale, which is dominated by lattice
artefacts, that is, the largely arbitrary details of the regularized set-up at this scale. On the other hand,
the large gravitational Wilson loops we have studied do not obviously contain retrievable 
curvature or other geometric information, unless it is hidden in higher-order correlators. Here one
should of course keep in mind that even in the classical continuum theory we do not know how to relate 
the values of non-infinitesimal Wilson loops on a general Riemannian manifold to its curvature,
because of the nonabelian nature of the metric connection. The most obvious quantities to try to
define and investigate are therefore lattice Wilson loops which are much bigger than the minimal loops,
but sufficiently small to have an interpretation in terms of a suitably averaged curvature, in the
continuum limit. The challenge is at least two-fold: (i) to define suitable classes of closed curves which
have an invariant meaning when we integrate over all geometries, and (ii) to use them to find a notion of 
(quantum) curvature whose expectation value
remains finite and well-defined at the Planck scale, while converging to some  
function of the continuum Riemann tensor in the classical limit.  
The Regge definition of (scalar) curvature on piecewise linear geometries is simple, but offers little insight 
into how to define a well-behaved notion of quantum curvature. 
The use of holonomies may offer an alternative, more flexible tool to achieving this goal. 
Further research into this issue is currently under way.

\subsection*{Acknowledgments} 
JA and AG  acknowledge support from the ERC Advanced Grant 291092
``Exploring the Quantum Universe'' (EQU) and by FNU, 
the Free Danish Research Council, through the grant 
``Quantum Gravity and the Role of Black Holes''. JJ acknowledges the 
support of grant DEC-2012/06/A/ST2/00389 from 
the National Science Centre Poland. 
The contribution of RL is part of the 
research programme of the Foundation for Fundamental Research 
on Matter (FOM), financially supported by the Netherlands 
Organisation for Scientific Research (NWO).

\end{document}